\documentclass[aps,prL,twocolumn,superscriptaddress,showkeys]{revtex4-1}
\usepackage{graphicx}
\usepackage{dcolumn}
\usepackage{amsmath}
\usepackage[english]{babel}
\usepackage[utf8]{inputenc}
\begin{document}
	\title{Tunable coupling and isolation of single electrons in silicon metal-oxide-semiconductor quantum dots}
	
	\author{H.G.J. Eenink}
	\thanks{These authors contributed equally to this work.}
	\affiliation{QuTech and Kavli Institute of Nanoscience, TU Delft, P.O. Box 5046, 2600 GA Delft, Netherlands}
	\author{L. Petit}
	\thanks{These authors contributed equally to this work.}
	\affiliation{QuTech and Kavli Institute of Nanoscience, TU Delft, P.O. Box 5046, 2600 GA Delft, Netherlands}
	\author{W.I.L. Lawrie}
	\affiliation{QuTech and Kavli Institute of Nanoscience, TU Delft, P.O. Box 5046, 2600 GA Delft, Netherlands}
	\author{J.S. Clarke}
	\affiliation{Components Research, Intel Corporation, 2501 NE Century Blvd, Hillsboro, Oregon 97124, USA}
	\author{L.M.K. Vandersypen}
	\affiliation{QuTech and Kavli Institute of Nanoscience, TU Delft, P.O. Box 5046, 2600 GA Delft, Netherlands}
	\author{M.Veldhorst}
	\email{corresponding author: m.veldhorst@tudelft.nl}
	\affiliation{QuTech and Kavli Institute of Nanoscience, TU Delft, P.O. Box 5046, 2600 GA Delft, Netherlands}
	\pacs{}
	\keywords{Silicon, tunnel coupling, quantum dots}
	\begin{abstract}
		Extremely long coherence times, excellent single-qubit gate fidelities and two-qubit logic have been demonstrated with silicon metal-oxide-semiconductor spin qubits, making it one of the leading platforms for quantum information processing. Despite this, a long-standing challenge in this system has been the demonstration of tunable tunnel coupling between single electrons. Here we overcome this hurdle with gate-defined quantum dots and show couplings that can be tuned on and off for quantum operations. We use charge sensing to discriminate between the (2,0) and (1,1) charge states of a double quantum dot and show excellent charge sensitivity. We demonstrate tunable coupling up to 13 GHz, obtained by fitting charge polarization lines, and tunable tunnel rates down to below 1 Hz, deduced from the random telegraph signal. The demonstration of tunable coupling between single electrons in a silicon metal-oxide-semiconductor device provides significant scope for high-fidelity two-qubit logic toward quantum information processing with standard manufacturing.
	\end{abstract}
	\maketitle
	
	Quantum computation with quantum dots has been proposed using qubits defined on the spin states of one \cite{loss1998quantum}, two \cite{levy2002universal} or more \cite{divincenzo2000universal,shi2012fast} electrons. In all these proposals, a crucial element required to realize a universal quantum gate set is the exchange interaction between electrons. The exchange interaction is set by the tunnel coupling and the detuning, and gaining precise control over these parameters enables to define and operate qubits at their optimal points \cite{martins2016noise,reed2016reduced,taylor2013electrically, kim2014quantum}. Excellent control has already been reported in GaAs \cite{hensgens2017quantum, reed2016reduced, martins2016noise}, strained silicon \cite{borselli2015undoped, zajac2018resonantly} and more recently in strained germanium \cite{hendrickx2018gate,hendrickx2019fast}. Reaching this level of control in silicon metal-oxide-semiconductor (SiMOS) quantum dots is highly desired as this platform has a high potential for complete integration with classical manufacturing technology \cite{sabbagh2019quantum, mazzocchi201999, maurand2016cmos}. This becomes apparent from many proposals of architectures for large-scale quantum computation \cite{loss1998quantum,veldhorst2017silicon, vandersypen2017interfacing, li2018crossbar, taylor2005fault, friesen2007efficient,trauzettel2007spin} that make use of full control over the exchange interaction. However, current two-qubit logic with single spins in SiMOS is based on controlling the exchange using the detuning only \cite{veldhorst2015two} or is executed at fixed exchange interaction \cite{huang2019fidelity}. 
	
	A first step toward the required control has been the demonstration of tunable coupling in a double quantum dot system operated in the many-electron regime, where gaining control is more accessible owing to the larger electron wave function \cite{tracy2010double,lai2011pauli}. More recently, exchange-controlled two-qubit operations have been shown with three-electron quantum dots \cite{yang2019silicon}. However, tunnel couplings between single electrons that can be switched off and turned on for qubit operation still remain to be shown in SiMOS.
	
	In this work we show a high degree of control over the tunnel coupling of single electrons residing in two gate-defined quantum dots in a SiMOS device. The system is stable and no unintentional quantum dots are observed. We are able to measure charge transitions using a sensitive single-electron-transistor (SET) as charge sensor and characterize the system in the single-electron regime. From a comparison of charge stability diagrams of weakly and strongly coupled double quantum dots, we conclude that we control the tunnel coupling by changing quantum dot location. We show that we can effectively decouple the double quantum dot from its reservoir and control the inter-dot tunnel coupling of the isolated system with a dedicated barrier gate. We quantify the tunability of the coupling by analyzing charge polarisation lines and random telegraph signals, and find tunnel couplings up to 13 GHz and tunnel rates down to below 1 Hz. 
	\section{Results and discussion}
	\begin{figure*}%
		\includegraphics[width=504 pt]{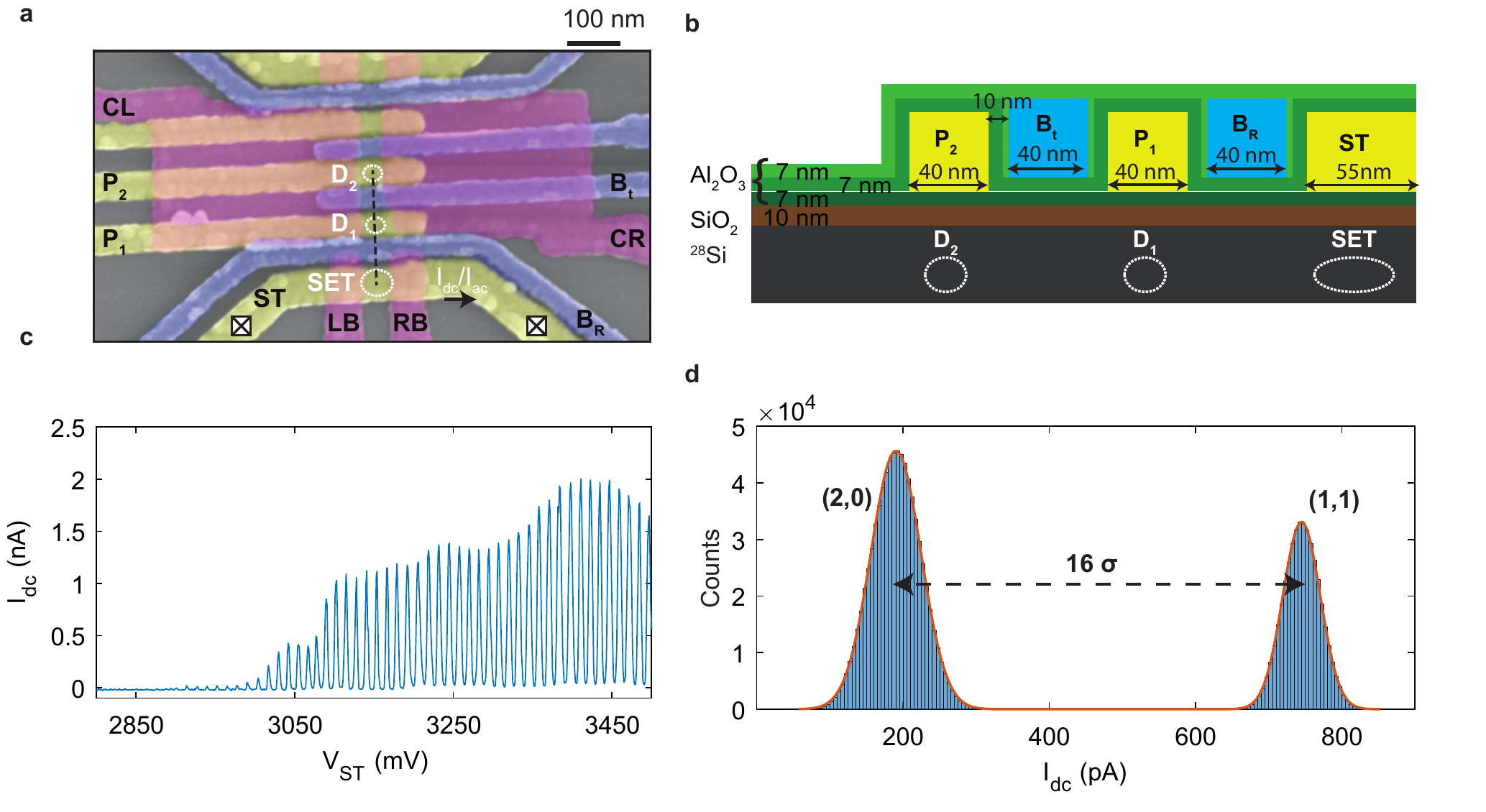}%
		\caption{Device layout and SET characterisation. \textbf{a} False-colour scanning electron micrograph (SEM) of a device identical to the one measured. Purple, yellow and blue colourings correspond to the first, second and third metal layers respectively. Crossed boxes indicate the ohmic source and drain contacts used to measure I\textsubscript{dc} and I\textsubscript{ac}, circles indicate the intended location of the quantum dots D\textsubscript{1} and D\textsubscript{2} and the single-electron-transistor (SET). The quantum dots are defined using gate electrodes P$_1$ and P$_2$, confined laterally using CL and CR. B$_t$ controls the tunnel coupling between the quantum dots and B$_R$ the tunnel coupling to the SET. \textbf{b} Schematic of a cross-section of the device along the quantum dot region (dashed line in \textbf{a}), indicating key dimensions and dot locations. \textbf{c} Transport source-drain current I$_{dc}$ versus top gate voltage V$_{ST}$ of the SET defined using gate electrodes ST, LB and RB. Regular spacing of Coulomb peaks indicates a well defined quantum dot, ideal for charge sensing. \textbf{d} Histogram of the charge sensor current as a response to (2,0)-(1,1) tunneling events. The counts are extracted from 4655 single-shot traces with integration time t$_i$= 82 $\mu$s, measurement bandwidth 0-50 kHz, and bin size b = 5 pA. The peaks are fitted with a double Gaussian with $\sigma_{(2,0)}$ = 34.1 pA and $\sigma_{(1,1)}$ = 25.5 pA, giving a peak spacing of over 16 $\sigma_{(2,0)}$.}
		\label{fig:1}
	\end{figure*}
	
	Figure 1a shows a scanning electron micrograph (SEM) of a SiMOS device nominally identical to the one measured and Fig. 1b shows a schematic cross-section of the quantum dot region along the dashed line in Fig. 1a. A high quality wafer is realized \cite{sabbagh2019quantum} with a 100 nm $^{28}$Si epilayer with an 800 ppm residual $^{29}$Si concentration \cite{veldhorst2014addressable}, covered by 10 nm thermally grown SiO$_2$. Ohmic contacts are made by defining highly doped n$^{++}$ regions by phosphorus-ion implantation. We use an overlapping gate integration scheme \cite{angus2007gate,borselli2015undoped,zajac2016scalable} and use palladium (Pd) gates, which have the beneficial property of small grain size \cite{brauns2018palladium}. The gates are electrically isolated by an Al$_{2}$O$_{3}$ layer grown by atomic layer deposition. The sample is annealed at 400 $^{\circ}$C in a hydrogen atmosphere to repair e-beam induced damage to the silicon oxide and to reduce the charge trap density \cite{kim2017annealing, nordberg2009enhancement}.
	
	Figure 1c shows the current through the SET, electrostatically defined using gates ST, LB and RB, that is used as charge sensor and as an electron reservoir. The highly regular coulomb peak spacing indicates a well defined quantum dot, which has a constant charging energy of approximately 0.9 meV. We extract a gate capacitance of 13 aF, in agreement with a simple parallel plate capacitor model. We form a double quantum dot between the confinement barriers CL and CR, using the gates P\textsubscript{1} and P\textsubscript{2} to tune the quantum dot potentials. B\textsubscript{t} and B\textsubscript{R} are used to control the tunnel coupling between the quantum dots and from the quantum dots to the SET, respectively.
	
	We characterize the charge readout sensitivity by recording the random telegraph signal (RTS) originating from the tunneling of the electrons between the (2,0) and (1,1) charge states with $\Gamma_c \approx$ 48 Hz, with $\Gamma_{c}$ being the inter-dot tunnel rate.
	The fidelity of the (2,0)-(1,1) charge readout is often limited by the sensitivity of the charge sensor to inter-dot transitions. We have designed and positioned the SET with respect to the double quantum dot in such a way that this sensitivity is maximized. 
	Figure 1d shows a histogram of the measured readout signal, using an integration time $\tau$ = 82 $\mu$s. We fit the counts with a double Gaussian curve with $\mu_{(2,0), (1,1)}$ and $\sigma_{(2,0), (1,1)}$ the mean and standard deviation of the Gaussian distributions corresponding to the two charge states. We find $\Delta\mu_{(2,0)-(1,1)}$ $>$ 16 $\sigma_{(2,0)}$ corresponding to an excellent discrimination between the (2,0) and (1,1) charge states. 
	
	\begin{figure*}%
		\includegraphics[width=\linewidth]{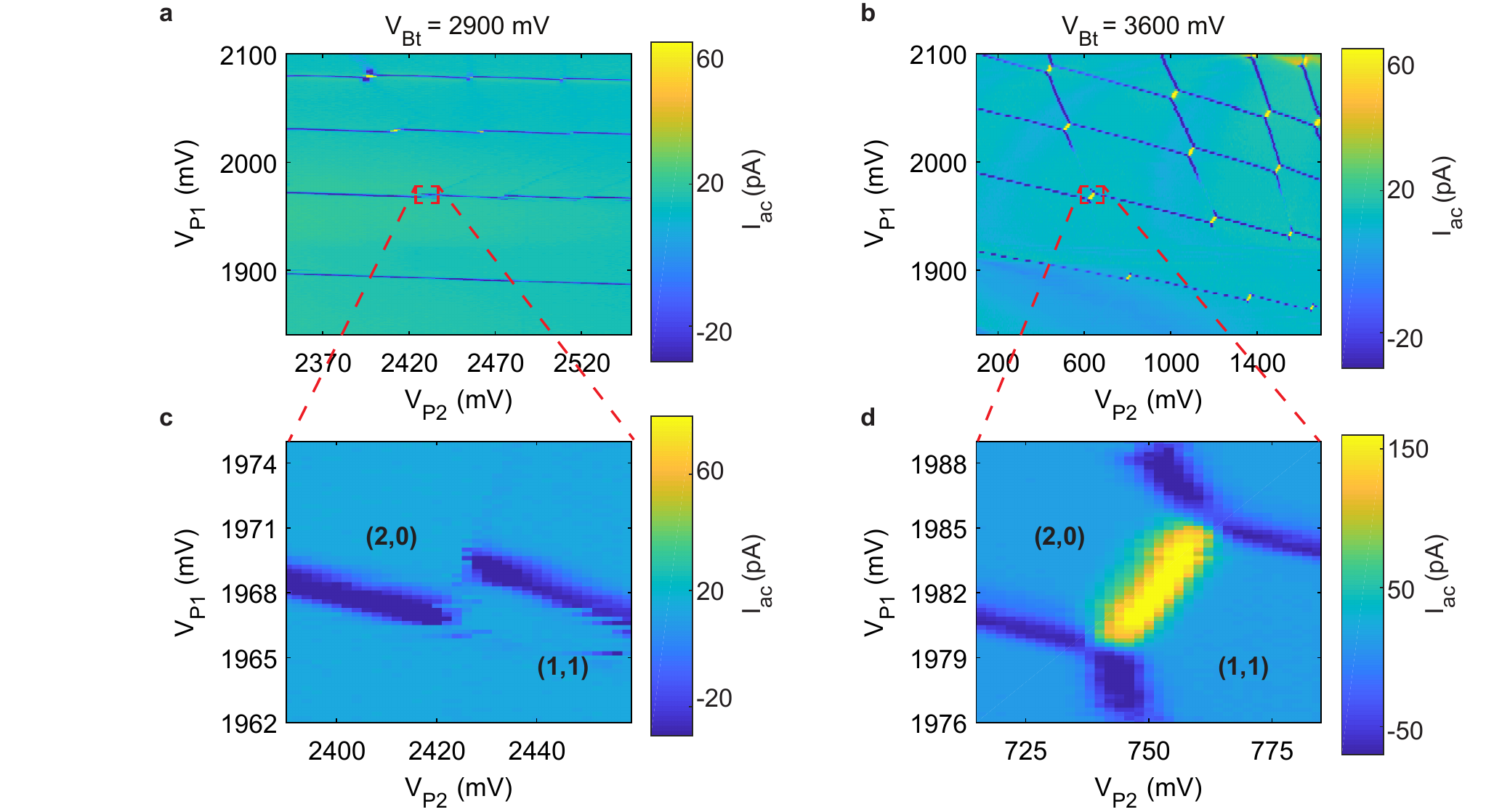}%
		\caption{Double quantum dot charge stability diagrams. \textbf{a}, \textbf{b} Charge stability diagrams of the charge sensor response I$_{ac}$ as a function of voltages V$_{P2}$ and V$_{P1}$ of a double quantum dot for weak (\textbf{a}, V$_{Bt}$ = 2.9 V) and strong (\textbf{b}, V$_{Bt}$ = 3.6 V) coupling. Electrons are loaded from the SET. Transitions with a tunnel rate $\Gamma < f_{ac}$ are not visible. \textbf{c}, \textbf{d} High resolution zoom in of the (2,0)-(1,1) anticrossing for both weak (\textbf{c}) and strong (\textbf{d}) tunnel coupling.}
		\label{fig:2}
	\end{figure*}
	
	To precisely measure charge transitions, we implement charge sensing using a lock-in amplifier and apply a square wave excitation at $f_{ac}$ = 77 Hz on the gate B\textsubscript{t}.
	Figure 2a and 2b show the double quantum dot charge stability diagrams of the charge sensor response as a function of V$_{P2}$ and V$_{P1}$ for weak (V$_{Bt}$ = 2.9 V) and strong (V$_{Bt}$ = 3.6 V) coupling.
	Horizontal and vertical blue lines indicate the loading of an additional electron from the SET to quantum dots D\textsubscript{1} (located under the gate P\textsubscript{1}) and D\textsubscript{2} (located under P\textsubscript{2}) respectively, while diagonal yellow lines indicate electron transitions between the two quantum dots. We do not observe more charge transitions at voltages lower than the measured range (see Supplemental Figure 1) and we conclude that the double quantum dot is in the single electron regime. In order to highlight the difference between weak and strong coupling, Fig. 2c and 2d show higher resolution maps of the (2,0)-(1,1) anticrossing. 
	
	When we set a weak inter-dot coupling, charge addition lines of D\textsubscript{2} are barely visible in the charge stability diagram, because of the low tunnel rate between D\textsubscript{2} and the reservoir. This indicates that the tunnel rate is significantly smaller than the excitation frequency applied to the gate. Similarly, at the (2,0)-(1,1) inter-dot transition, no transitions between the quantum dots can be observed because of the low inter-dot coupling. The loading of the first electron in  D\textsubscript{2} can only be observed from the shift of the  D\textsubscript{1} charge addition line, caused by the mutual capacitance $E_m$ of the two quantum dots. Only in the multi-electron regime where the quantum dot wave functions are larger and have more overlap, the coupling is sufficiently high to observe charge transition lines.
	
	When the inter-dot coupling is strong, charge addition lines belonging to D\textsubscript{2} are visible near the anticrossings and at high V$_{P1}$, where $\Gamma_{R2}$ is increased. Additionally, $t_c$ and $E_m$ are increased and we observe a honeycomb shaped charge stability diagram, with clearly visible inter-dot transition lines, even when only a single electron is loaded on each quantum dot. 
	
	We estimate the relative location and size of the quantum dots from the gate voltage differences $\Delta$V$_{P1(2)}$ needed to load the second electron with respect to the first electron. We additionally use the cross-capacitances $\alpha_{r1(2)}$ of the plunger gates, determined by measuring the shift in V$_{P1(2)}$ of the charge transition line of the first electron in D\textsubscript{1(2)} as a function of a step in V$_{P2(1)}$, where $\alpha_{r1(2)}$ is the ratio between the shift and the step.
	
	When the coupling is weak, we find $\Delta$V$_{P1}$ $\approx$ 70 mV, $\alpha_{r1} <$ 0.05 for D\textsubscript{1} and $\Delta$V$_{P2}$ $\approx$ 50 mV, $\alpha_{r2} \approx$ 0.33 for D\textsubscript{2}. We conclude that we have a system of two weakly coupled quantum dots located under P\textsubscript{1} and P\textsubscript{2}.
	
	We now analyse how the locations of D\textsubscript{1} and D\textsubscript{2} change from the changes in $\Delta$V$_{P}$ and $\alpha_r$. For D\textsubscript{1}, both $\Delta$V$_{P1}$ and $\alpha_{r1}$ are almost independent of the coupling. 
	For D\textsubscript{2}, $\Delta$V$_{P2}$ increases by a factor 11, from $\Delta$V$_{P2}\approx$ 50 mV for weak coupling to $\Delta$V$_{P2}\approx$ 550 mV for strong coupling, while $\alpha_{r2}$ increases by a factor 5, from 0.3 to 1.5. 
	The increase in $\alpha_{r2}$ can be explained by a change in the location of D\textsubscript{2} toward the gate P\textsubscript{1}, to a position partly below the gate B\textsubscript{t}. This change of quantum dot location will decrease the lever arm and this is likely the cause of the increase in $\Delta$V$_{P2}$. We conclude that tuning from weak to strong coupling causes the location of D\textsubscript{2} to shift from a position mostly under P\textsubscript{2} to a position partly below B\textsubscript{t}, while D\textsubscript{1} is stationary under P\textsubscript{1}. The ease with which D\textsubscript{2} can be displaced additionally suggests that no unintentional quantum dots are formed between barrier gates.
	
	\begin{figure*}
		\includegraphics[width=504 pt]{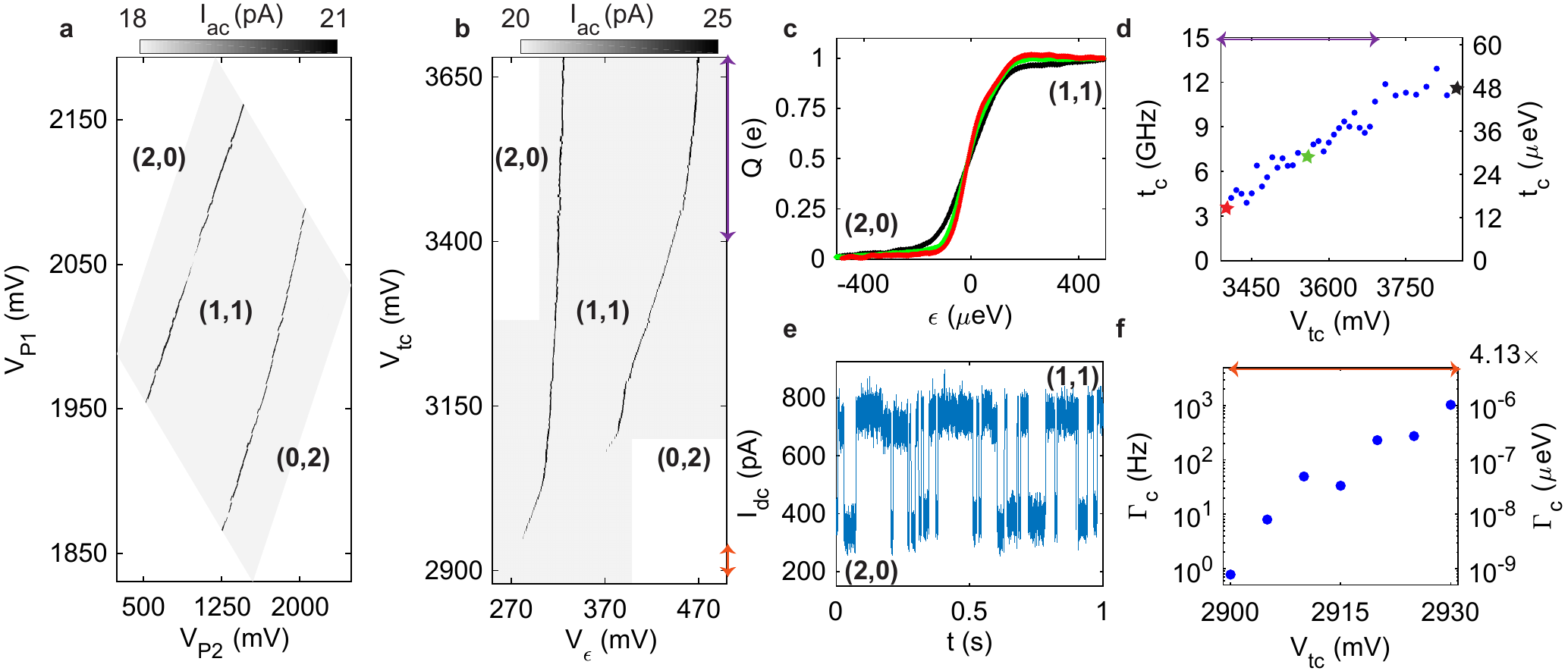}%
		\caption{Charge stability diagrams and tunnel coupling control of an isolated double quantum dot. \textbf{a} Map of the isolated (2,0)-(1,1) and (1,1)-(0,2) anticrossings as a function of V$_{P2}$ and V$_{P1}$. No additional electrons are loaded into the quantum dot islands due to a negligible $\Gamma_R$. \textbf{b} Map of the (2,0)-(1,1) and (1,1)-(0,2) anticrossings as a function of detuning and barrier voltage. The relative lever arm between V$_{tc}$ and V$_\epsilon$ changes at lower barrier voltages, due to a change in quantum dot location. The orange and purple arrows indicate the ranges in which the tunnel coupling was determined using RTS and polarisation line measurements respectively. \textbf{c} Polarization lines (excess charge $Q$ as a function of detuning $\epsilon$) across the anticrossing for high $t_c$ (black, V$_{tc}$ = 3.85 V), intermediate $t_c$ (green, V$_{tc}$  = 3.6 V) and relatively low $t_c$ (red, V$_{tc}$  = 3.4 V). \textbf{d} Extracted $t_c$ from polarization lines as a function of V$_{tc}$, where we find tunable $t_c$ up to 13 GHz. \textbf{e} RTS for weak coupling V$_{tc}$ = 2.910 V. \textbf{f} Extracted $\Gamma_c$ from RTS measurements as a function of V$_{tc}$, demonstrating tunable tunnel rates down to below 1Hz.}
		\label{fig:3}
	\end{figure*}
	
	By reducing V$_{BR}$, the tunnel rate $\Gamma_R$ between the the SET reservoir and the quantum dots can be reduced and the loading and unloading of electrons can be prevented, resulting in an isolated quantum dot system \cite{bertrand2015quantum,yang2019silicon}. Because the reservoir is connected to room temperature electronics, decoupling the quantum dot from it may provide the advantage of reduced noise \cite{rossi2012electron}. Figure 3a shows the (2,0)-(1,1) and (1,1)-(0,2) anticrossings as a function of V$_{P2}$ and V$_{P1}$ when the coupling is strong. Only inter-dot transition lines are present over a wide range of voltages, much larger than the $\Delta$V$_{P}$ extracted in the previous section. This implies that no additional electrons are loaded, as a result of a negligible coupling to the reservoir.
	The ability to control the inter-dot transitions of a double quantum dot without loading additional electrons provides good prospects for the operation of quantum dot arrays that are only remotely coupled to reservoirs, as proposed in quantum information architectures \cite{veldhorst2017silicon,li2018crossbar,taylor2005fault}.
	
	We control the tunnel coupling $t_{c}$ with the gate B\textsubscript{T}. To compensate for the influence of V$_{Bt}$ on detuning $\epsilon$ and on-site potential $U$, we implement virtual gates using a cross-capacitance matrix \cite{baart2016single,hensgens2017quantum,mills2019shuttling} and convert V$_{P2}$, V$_{P1}$ and V$_{Bt}$ to $\epsilon$, $U$ and t$_{c}$. Figure 3b shows the (2,0)-(1,1) and (1,1)-(0,2) anticrossings as a function of the new set of virtual gates V$_{\epsilon}$ and V$_{tc}$. For both transitions the inter-dot line vanishes at low V$_{tc}$, meaning that the coupling has been largely switched off. We observe that for the (1,1)-(0,2) anticrossing, the transition line disappears at V$_{tc} <$ 3.1 V, while for the (2,0)-(1,1) anticrossing this happens for V$_{tc} <$ 2.95 V. The variation may come from a small asymmetry in the system.
	
	We tune the double quantum dot to a significantly coupled regime and quantitatively analyze the system by taking charge polarization lines. Figure 3c shows charge polarization lines at high, intermediate and relatively low tunnel couplings within this regime. We measure the charge sensor response $Q$ as a function of detuning $\epsilon$ and fit the data according to a two level model that includes cross-talk of $\epsilon$ to the charge sensor and the influence of the quantum dot charge state on the charge sensor sensitivity \cite{dicarlo2004differential, hensgens2017quantum}. From the thermal broadening of the polarization line at low tunnel coupling, we extract the lever arm of V$_{\epsilon}$ for the detuning axis $\alpha_{\epsilon} \approx$ 0.04 eV/mV, by assuming the electron temperature to be equal to the fridge temperature of 0.44 K.
	
	For relatively low tunnel couplings, we observe in the charge polarization lines deviations from the model for a two-level system \cite{dicarlo2004differential} (see the red curve in Fig. 3c with $\epsilon >$ 0). This deviation can also not be explained by a modified model that includes valley states, considering an adiabatic detuning sweep and assuming zero temperature \cite{zhao2018coherent}. While these measurements were done adiabatically, the elevated temperature of 0.44 K can cause a non-negligible population of valley or other excited states. These excited states can cause a charge transition at a different detuning energy, thereby giving rise to a deviation. A large tunnel coupling can increase the relaxation rate of these excited states and thus decrease their population. As a consequence, the charge polarization lines are in agreement with the model for a two-level system \cite{dicarlo2004differential} at larger tunnel couplings.
	
	At tunnel couplings below 3 GHz, the thermal broadening of the polarization line prevents accurate fitting. Instead of the tunnel coupling energy $t_c$ we determine the inter-dot tunnel rate $\Gamma_c$, which is proportional to the square of the tunnel coupling \cite{braakman2013long, gamble2012two, korotkov1999continuous}. We measure the RTS (Fig. 3e) at the (2,0)-(1,1) transition and fit the counts $C$ of a histogram of the tunnel times $T$ to $C = Ae^{-\Gamma_c T}$, where $A$ is a normalisation constant. In the measurements we have tuned V$_\epsilon$ such that $\Gamma_{c(2,0)-(1,1)}\approx\Gamma_{c(1,1)-(2,0)}$.
	
	Figure 3d shows a $t_c$ as a function of V$_{tc}$, demonstrating tunable tunnel coupling in the strong coupling regime and Fig. 3f shows the obtained $\Gamma_c$ as a function of V$_{tc}$ from 1 kHz down to below 1 Hz. We note that we can further reduce the tunnel rate to even smaller rates simply by further reducing V\textsubscript{tc}.
	
	A change in barrier height or width results in an exponential change in t\textsubscript{c} and in $\Gamma_c$. When the tunnel coupling is low, D\textsubscript{2} is located mainly under P\textsubscript{2}, and a change in V$_{tc}$ has a significant impact on the barrier. Correspondingly, we observe an exponential dependence of $\Gamma_c$ versus V$_{tc}$. When the tunnel coupling is high, D\textsubscript{2} is located mostly under B\textsubscript{t} and the impact of V$_{tc}$ on the barrier is vanishing. As a result we observe a seemingly linear dependence of t\textsubscript{c} versus V$_{tc}$ from 3 up to 11 GHz that saturates around 13 GHz for V$_{tc} >$ 3675 mV.
	
	\section{Conclusions}
	We have demonstrated control over the tunnel coupling of single electrons residing in a double quantum dot in SiMOS. The inter-dot coupling of the (2,0)-(1,1) charge transition can be controlled by a barrier gate which changes the quantum dot location. We have demonstrated control over the tunnel coupling in the strong coupling regime from 3 to 13 GHz, as well as control over the tunnel rate in the weak coupling regime from 1 kHz to below 1 Hz. Achieving this degree of control in an isolated system constitutes a crucial step toward independent control over detuning and tunnel coupling for operation at the charge symmetry point \cite{martins2016noise,reed2016reduced}, and reaching the control required for large-scale quantum computation with quantum dots \cite{loss1998quantum,veldhorst2017silicon, vandersypen2017interfacing, li2018crossbar, taylor2005fault, friesen2007efficient,trauzettel2007spin}. While SiMOS systems are often said to be severely limited by disorder, the excellent control shown here provides great prospects to operate larger arrays fabricated using conventional semiconductor technology.
	
	\section{Acknowledgements}
	We thank Max Russ for helpful discussions. H.G.J.E, L.P and M.V. are funded by a Netherlands Organization of Scientific Research (NWO) VIDI grant. Research was sponsored by the Army Research Office (ARO) and was accomplished under Grant No. W911NF- 17-1-0274. The views and conclusions contained in this document are those of the authors and should not be interpreted as representing the official policies, either expressed or implied, of the Army Research Office (ARO), or the U.S. Government. The U.S. Government is authorized to reproduce and distribute reprints for Government purposes notwithstanding any copyright notation herein.
	
\providecommand{\noopsort}[1]{}\providecommand{\singleletter}[1]{#1}%
\providecommand{\refin}[1]{\\ \textbf{Referenced in:} #1}

\end{document}